\documentstyle[11pt,epsf,twocolumn]{article}
\begin{document}
\begin{titlepage}
\begin{center}
{\Huge
Meson Regge trajectories in relativistic  quantum
mechanics\\
}
$$
$$
$$
$$
{\large
{\bf V.V.Andreev}
{\it Gomel State
University, Gomel 246699, Belarus } \\
\medskip
{\bf M.N.Sergeenko}\\
\medskip
{\it Institute of Physics NASB, Minsk 220072, Belarus
and \\
Department of Physics, University of Illinois,  Chicago,
Illinois 60607, USA }
}
$$
$$
{\large
To published in Proceedings
8 Annual Seminar "Nonlinear Phenomena In Complex System" (NPCS'99)\\
(Minsk, Belarus, 1999) \\ }
\end{center}
$$
$$
$$
$$
\begin{center}
{\Huge Abstract}
\end{center}

{\Large
\begin{center}
A model potential for two-particle relativistic systems is investigated in
the framework of Poincare-invariant quantum mechanics (or relativistic
Hamiltonian dynamics). The potential considered allows to reduce the main
integro-differential equation of Poincare-invariant quantum mechanics to the
equation analogous to the radial equation and have analytical solution for
relativistic bound system. We discuss the possible choice of the parameters of
potential and apply our model to the description of light meson
Regge trajectories.
\end{center}
}

\end{titlepage}
\newpage

\twocolumn[
\begin{center}
{\large \bf {\ Meson Regge trajectories in relativistic  quantum
mechanics } \\}
{ {\ V. V. Andreev \\{\it {\ Gomel State
University, Gomel 246699, Belarus }}} \\\medskip M. N. Sergeenko
\\\medskip {\it {\ Institute of Physics NASB, Minsk 220072, Belarus
and \\Department of Physics, University of Illinois,  Chicago,
Illinois 60607, USA }} } \end{center}

\begin{abstract}
A model potential for two-particle relativistic systems is investigated in
the framework of Poincare-invariant quantum mechanics (or relativistic
Hamiltonian dynamics). The potential considered allows to reduce the main
integro-differential equation of Poincare-invariant quantum mechanics to the
equation analogous to the radial equation and have analytical solution for
relativistic bound system.
We discuss the possible choice of the parameters of
potential and apply our model to the description of light
meson Regge trajectories.
\end{abstract}
\vspace{10mm}]
\section{\bf Introduction}

Relativistic few body problem has received a great attention in hadronic and
nuclear physics. The most complete results here exist for the case of two
particles. Description of the bound system in the relativistic quantum field
theory is founded on the four-dimensional covariant Bethe-Salpeter equation
\cite{BeteS}. However, this equation gives series of difficulties when the
practical calculations are made.

There exist various reductions of the two-body Bethe-Salpeter equation.
Different forms of this reduction were discussed Logunov and Tavkhelidze
\cite{LogTav}, Kadyshevsky \cite{Kad1}, Todorov \cite{Tod}, Gross \cite
{Gross}, Poluzou, Keister and Lev \cite{Keist1}-\cite{Lev4} and many others.

The relativistic two-body systems were analyzed by Weinberg \cite{Wein},
Frankfurt and Strikman \cite{FrS}, Kondratyuk and Terent'ev \cite{KonTer} in
infinite-momentum frame. Some authors use diagrammatic approach i.e. they
select leading diagrams and project them onto the three-dimensional space.
Others make use effective Hamiltonians, Faddev equation. Different forms of
the quasi-potential equation can be derived using this approach \cite{Kad2}-%
\cite{Kad3}.

In this work we apply the point form of relativistic quantum mechanics (RQM)
\cite{Keist1}-\cite{Lev4} to the description of meson Regge trajectories.
RQM is also known in the literature as relativistic Hamiltonian dynamics or
Poincare-invariant a quantum mechanics with direct interaction.

\section{\bf RQM formalism for quark-antiquark bound states}

The formulation of relativistic quantum mechanics differs from
non-relativistic quantum mechanics by the replacement of invariance under
Galilean transformations with invariance under Poincare transformations. The
dynamics of many-particle system in the RQM is specified by expressing the
ten generators of the Poincare group $\hat M_{\mu\nu}$ and $\hat P_\mu$ in
terms of dynamical variables. In the constructing generators for interacting
systems it is customary to start with the generators of the corresponding
non-interacting system (we shall write this operators without ''{\bf a hat}%
'') and then add interaction in the a way that is consistent with Poincare
algebra. In the relativistic case it is necessary to add an interaction $V$
to more than one generator in order to satisfy the commutation relations of
the Poincare algebra. Dirac \cite{Di1} observed, that there is no unique way
of separating the generators into dynamical subset (the generators including
interaction $V$) and kinematic subset. Kinematic subset must be
associated with some subgroup of Poincare group, usually called stability
group \cite{Leut} or kinematic subgroup. Thus, the construction for two
interacting particles proceeds as follows:

1. The two-particle Hilbert space of non-interacting system is defined as
tensor product of two one-particle Hilbert spaces. A two-body unitary
representation of Poincare group on the two-particle Hilbert space is
reducible. A basis in this space can be constructed from single-particle
bases:

\begin{equation}
\left| p_1\lambda _1\right\rangle \left| p_2\lambda _2\right\rangle \equiv
\left| m_1s_1;p_1\lambda _1\right\rangle \otimes \left| m_2s_2;p_2\lambda
_2\right\rangle  \label{bas1}
\end{equation}
with normalization
$$
\left\langle p_1^{\prime }\lambda _1^{\prime }\right| \left\langle
p_2^{\prime }\lambda _2^{\prime }\right| \left| p_1\lambda _1\right\rangle
\left| p_2\lambda _2\right\rangle
$$
\[
=\delta _{\lambda _1^{\prime }\lambda
_1}\delta _{\lambda _2^{\prime }\lambda _2}\delta (\vec p_1^{\,\prime} -
\vec p_1) \delta (\vec p_2^{\,\prime} - \vec p_2),
\]
where $p_i,m_i,s_i,\lambda _i$ are momenta, masses, spins and projects of
spin of particles, which will form bound system.

2. Clebsch-Gordan coefficients for Poincare group are constructed and used
to reduce the unitary representation on the two-particle Hilbert space to
linear superposition (direct integral) of irreducible representations.
Poincare generators for irreducible representations of the non-interacting
system are constructed, along with operators for the mass and spin and other
operators. The result of this step is
\[
\left| \vec P_{12},\mu ,\hskip 2pt\left[ J,k\right]
,(ls);[m_1s_1;m_2s_2]\right\rangle =
\]
\[
\sum_{ls}\sum_{\lambda
_1\lambda_2}\int d^3 k\sqrt{\frac{\omega_{m_1}\left( \vec p_1\right)
\omega_{m_2}\left(\vec p_2\right) M_0} {\omega_{m_1}(\vec k)
\omega_{m_2}(\vec k) \omega_{M_0}( \vec P_{12})}}
\]
\[
\sum_{m\lambda }\sum_{\nu_1\nu_2}\left\langle s_1\nu _1,s_2\nu _2\right|
\left. s\lambda \right\rangle \left\langle lm,s\lambda \right| \left. J\mu
\right\rangle Y_{lm}\left( \theta ,\phi \right)
\]
\begin{equation}
D_{\lambda_1\nu_1}^{1/2}\left( \vec
n\left( p_1,P_{12}\right) \right) D_{\lambda_2\nu _2}^{1/2}\left( \vec
n\left( p_2,P_{12}\right) \right)
\left| p_1\lambda _1\right\rangle \left| p_2\lambda_2\right \rangle ,
\label{state}
\end{equation}
where $\left\langle s_1\nu _1,s_2\nu _2\right| \left. s\lambda \right\rangle
$, $\left\langle lm,s\lambda \right| \left. J\mu \right\rangle $ are
Clebsh-Gordan coefficients of $SU(2)$-group, $Y_{lm}(\theta ,\phi )$ are the
spherical harmonics with spherical angle of $\vec k$. Also, in Eq. (\ref
{state}) $D^{1/2}\left( \vec n\right) = 1 -i\vec n\cdot\vec{\sigma } /\sqrt{%
1+\vec n^2}$ is $D$-function of Wigner rotation, which is determined by the
vector-parameter $\vec n(p_1,p_2) = \vec u_1\times \vec u_2 /(1 -\vec
u_1\cdot\vec u_2)$ with $\vec u = \vec p/\left(\omega_m\left(\vec p\right) +
m\right)$. The three-momenta $\vec p_1$, $\vec p_2$ of the particles (quarks
in our case) with the masses $m_1$ and $m_2$ of relativistic system are
transformed to the total, $\vec P_{12}$, and relative momenta $\vec k$ to
facilitate the separation of the center-of-mass motion:
\[
\vec P_{12} =\vec p_1+\vec p_2,
\]
\begin{equation}
\vec k = \vec p_1 + \frac{\vec P_{12}}{M_0} \left(\frac{\vec P_{12}\cdot\vec
p_1} {\omega_{M_0}(\vec P_{12}) + M_0} + \omega_{m_1}(\vec p_1)\right),
\label{veck}
\end{equation}
where
\begin{equation}
M_0 = \omega_{m_1}(\vec k) + \omega _{m_2}(\vec k)  \label{mass}
\end{equation}
mass of non-interacting system and $\omega_{m_1}(\vec p_1) = \sqrt{\vec
p_1^2 + m_1^2}$.

The state vector $\left| \vec P_{12},\hskip 2pt\mu ,\hskip 2pt\left[ J\hskip
2ptk\right] ,\hskip 2pt(l\hskip 2pts);[m_1s_1;m_2s_2]\right
\rangle$ $
\equiv \left| \vec P_{12},J,\mu ,k,\hskip 2pt(l\hskip
2pts)\right\rangle $ is the eigenstate of operators $\vec P_{12}$ , $J^2$, $
J_3$ and, also, $L^2$ and $S^2$, where $\vec L$ and $\vec S$ are relative
orbital momentum and total spin momentum, respectively. This vector of the
non-interacting $Q\bar q$ system transforms irreducibly under Poincare
transformations.

3. Following Bakamjian and Tomas, interactions are added to the mass
operator in the irreducible free-particle representation. The new mass
operator is defined by
\begin{equation}
\hat M\equiv M_0 +\hat V.  \label{massa}
\end{equation}
If $\hat V$ is any operator that satisfies the following conditions
\begin{equation}
\hat M = \hat M^{\dagger },\ \ \ M>0,  \label{cond1}
\end{equation}
\begin{equation}
[\vec P_{12},\hat V ]_{-} = [i\vec{\bigtriangledown}_{\vec p_{12}},\hat
V]_{-}=[ \vec J,\hat V]_{-}=0  \label{cond2}
\end{equation}
then the similar set of interacting particles will satisfy the same
commutation relations as the set of non-interacting system.

There are three forms of the dynamics in the relativistic quantum mechanics
called ''instant'', ''point'', and ''light-front'' forms \cite{Di1}. The
description in the instant form implies that the operators of three-momentum
and angular momentum do not depend on interactions, i.e. $\widehat{\vec P}$
= $\vec P$ and $\widehat{\vec J}$ = $\vec J~~(\vec J=(\hat M^{23},\hat
M^{31},\hat M^{12}$)) and interactions can be presented in terms of operator
$\hat P^0$ and generators of the Lorentz boosts $\vec N = (\hat M^{01},\hat
M^{02},\hat M^{03})$. The description in the point form implies that the
operators $\hat M^{\mu\nu }$ are the same as for non-interacting particles,
i.e. $\hat M^{\mu\nu } = M^{\mu\nu }$, and these interaction terms can be
presented only in the form of the four-momentum operators $\hat P$. In the
front form with the fixed $z$ axis we introduce the + and - components of
the four-vectors as $p^{+}=(p^0+p^z)/\sqrt{2}$, $p^{-}=(p^0-p^z)/\sqrt{2}$.
We require that in the front form the operators $\hat P^{+},\hat P^j,\hat
M^{12},\hat M^{+-},\hat M^{+j}$ $(j=1,2)$ are the same as the corresponding
free operators, and interaction terms can be presented via the operators $%
\hat M^{-j} $ and $\hat P^{-}$.

In our work we use the point form of RQM because:

\noindent
$\alpha$). The meson wave function (see below) being Lorentz invariant, as
(in a point form of dynamics) the operator of a boost does not contain
interaction. In the instant form of RQM this property of wave functions does
not take a place.

\noindent
$\beta$). The relativistic impulse approximation in an instant form and
dynamics on light front automatically breaks down the Poincare-invariance of
models. In the point form of RQM such a violation does not happen.

Given such an interaction $\hat V$ it is useful to define two other related
interactions \cite{Pol1},

\begin{equation}
\hat U = \hat M^2 - M_0^2 = \hat V^2 + M_0\hat V + \hat V M_0  \label{oper1}
\end{equation}

\begin{equation}
\hat{W} = \frac 14\left[(\hat M^2 - M_0^2) + (m_1^2-m_2^2)^2*\left(\frac
1{M^2} - \frac 1{M_0^2}\right)\right]  \label{oper2}
\end{equation}

The eigenvalue problem for the mass of $Q\bar q$ system can be expressed in
the three equivalent forms \cite{Pol1}:
\[
\hat M\mid \Psi >\equiv (M_0+\hat V)\mid \Psi > = M\mid \Psi >,
\]
\[
M_0^2 +\hat U\mid \Psi >=M^2\mid \Psi >,
\]
\begin{equation}
(k^2 + \hat W)\mid \Psi > = \eta \mid \Psi >,
\label{maineq3}
\end{equation}
where the mass $M$ of bound $Q\bar q$ system and $\eta $ have relationship,
\[
M^2 = 2\eta +m_1^2 + m_2^2+
\]
\begin{equation}
+2\sqrt{\eta (\eta +m_1^2+m_2^2) + m_1^2 m_2^2} .
\label{massa2}
\end{equation}

When $m_1=m_2=m$, the equation (\ref{massa2}) reduces to
\begin{equation}
M^2 = 4\left(\eta + m^2\right) .
\label{massa3}
\end{equation}

Solution of any of the above eigenvalue problems in the point form of RQM
leads to the eigenfunctions of the form
\[
\left\langle \vec V_{12},J,\mu ,k,(ls)\right. \left| \vec V,J^{\prime},\mu^{\prime}
,M\right\rangle =
\]
\begin{equation}
\delta_{JJ^{\prime }}\delta_{\mu\mu^{\prime }} \delta(\vec
V -\vec V_{12})\Psi^{J\mu}\left(k\hskip 2ptl\hskip 2pts\right)
\label{psi}
\end{equation}
with the velocities of the bound system $\vec V ={\vec P}/{M}$ and
non-interacting system $\vec V_{12} = {\vec P_{12}}/{M_0}$. The function $
\Psi^{J\mu }\left(k\hskip 2ptl\hskip 2pts\right)$ satisfies (in the point
form) the following equation \cite{Pol1},
\[
\sum_{l^{\prime }s^{\prime }}\int\limits_0^\infty <k\hskip 2pt(l\hskip 2pt
s)\parallel W^J\parallel k^{\prime }\hskip 2pt(l^{\prime }\hskip 2pt
s^{\prime })>\Psi ^J(k^{\prime }\hskip 2pt l^{\prime }\hskip 2pts^{\prime
})k^{\prime }{}^2dk^{\prime }+
\]
\begin{equation}
k^2\Psi ^J(k\hskip 2pt l\hskip 2pt s) =
\eta \Psi ^J(k\hskip 2pt l s)
\label{maineq}
\end{equation}
with reduced matrix element of operator $\hat W$
\[
\left\langle \vec V_{12},J,\mu ,k,(ls)\right| W\left| \vec V_{12}^{\prime
},J^{\prime },\mu ^{\prime },k^{\prime },(l^{\prime }s^{\prime
})\right\rangle =
\]
\begin{equation}
\delta _{JJ^{\prime }}\delta _{\mu \mu ^{\prime }}\delta (\vec V_{12}-\vec
V_{12}^{\prime })\left\langle k,(ls)\right| \left| W^J\right| \left|
k^{\prime },(l^{\prime }s^{\prime })\right\rangle .  \label{redmat}
\end{equation}
Equation (\ref{maineq}) is a radial equation. Therefore we have used
Poincare group properties to separate the radial and angular dependencies.

As vector of irreducible basis for a system with interaction and without its
satisfies to conditions of a normalization and completeness, the wave
functions satisfy to the following normalization condition,

\begin{equation}
\sum_{ls}\int_0^\infty dk\hskip 2ptk^2\left| \Psi ^J\left( kls\right)
\right| ^2=1.  \label{norm}
\end{equation}

\section{\bf Model potential and solving main equation.}

In this section we apply the formalism developed above to calculate mass
spectra of mesons containing $u$, $d$ and $s$ quarks. To choose appropriate
interquark potential we use a well-known experimental fact that light
hadrons populate approximately linear Regge trajectories, i.e., $%
M^2\simeq\beta J + const$, with the same slope, $\beta \simeq 1.2$ $GeV^2$,
for all trajectories (see, for example, \cite{Kon1}). Therefore, we can take
the effective model potential $W$ in the oscillator form

\begin{equation}
\hat W(r) = W_0 \hskip 2pt \delta \left(\vec r \right) + \beta^4 r^2,
\label{poten} \end{equation}
where $W_0$ and $\beta$ are free
parameters. As follows from calculations of meson spectra done by
Godfrey and Isgur \cite{Godfrey}, spin-dependent corrections are
important for $1S$, $2S$, $1^3P_2$ and $1^3P_2$ states.  However for
other states spin-dependent corrections are small and they can be
neglected in Eq. (\ref{poten}). Using standard relationships for
operator $\hat r$ and orbital momentum operator $\hat{\vec L^2}$
\[
\left\langle k\right | \hat r\left | k^{\prime}\right\rangle = -i\vec{%
\nabla_k}\delta (k-k^{\prime}),
\]
\[
\hat{\vec L^2}\left| \vec V_{12},J,\mu,k,\hskip
2pt(l\hskip 2pts)\right\rangle = l(l+1)\left| \vec V_{12},J,\mu,k,\hskip %
2pt(l\hskip 2pts)\right\rangle
\]
we reduce main integro-differential equation of RQM to the ordinary
quantum-mechanical radial equation with the oscillator potential (only
impulse representation),
\[
\left[\frac{d^2}{d k^2} + \frac 2k\frac d{d k} - \frac{l(l+1)}{k^2} -\frac{
k^2}{\beta^4}\right]\Psi\left(kls\right)=
\]
\begin{equation}
=\frac{W_0 -\eta}{\beta^4}\Psi\left(kls\right) .
\label{meq}
\end{equation}

The eigenfunctions of Eq. (\ref{meq}) are
\[
\Psi(kls) = N_{nl}\exp\left(-\frac{k^2} {2\beta^2}\right)*
\]
\begin{equation}
*\left(\frac{k}{\beta}\right)^{l} F\left(-n,l+\frac
32,\frac{k^2}{\beta^2}\right) \label{fun1} \end{equation}
with
\[
N_{nl}=\frac{\beta ^{-l-3/2}}{\Gamma (l+3/2)}\sqrt{\frac{\Gamma (n+l+3/2)}{
\Gamma (n+1)},}
\]
where $n,l=0,1,2\ldots $, $F(a,b,z)$ is the hypergeometric function, $\Gamma
(n)$ is the Gamma function. Note that the wave function of the ground state (
$n,l=0)$ has the Gaussian form, which is used in many relativistic models of
hadrons. This wave functions gives an excellent description of
electromagnetic formfactors at small transfers $Q^2$, $Q^2<$a few $GeV$.

Quantization condition is defined by
\begin{equation}
\eta = W_0+2\beta^2\left(2n+l+3/2\right) .  \label{quancon}
\end{equation}

The spectra of mesons composed of quarks with equal masses ($m_1=m_2\equiv m$
), for example, $u$ and $d$ quarks, are given by (see Eq. (\ref{massa3}) and
Eq. (\ref{quancon}) )
\begin{equation}
M^2 = 4(m^2+W_0) + 8\beta^2\left(2n+l+\frac 32\right) .  \label{spect}
\end{equation}
Thus we reproduce the linear dependence $M^2(l)$ in the framework of the
two-body relativistic equation (\ref{maineq}). The parameters $(m^2 + W_0)$
and $\beta $ have been found from the fitting the Regge trajectories (see
Fig.1),

\begin{figure}[thb]
\vspace*{-0.2cm}
\epsfxsize = 8cm
\epsffile{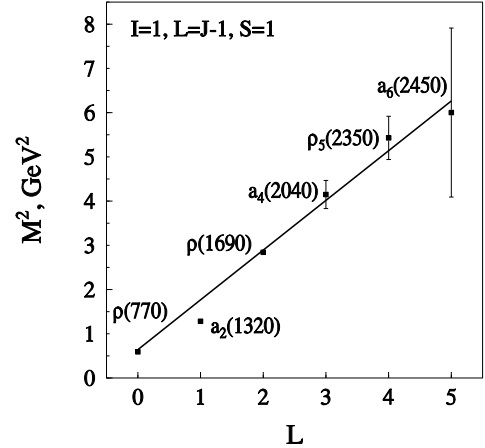}
\vspace*{-1.4cm}
\caption{$\rho-meson$ Regge trajectory}
\end{figure}

\[
(m^2+W_0) = -0.261\;GeV^2,\ \ \ \beta = 0.375\;GeV.
\]
If we take the same masses of $u$ and $d$ quark as in Ref. \cite{Godfrey},
i.e.,
\begin{equation}
m_u = m_d\equiv m = 0.22\;GeV,  \label{um}
\end{equation}
then we obtain
\[
W_0 = -0.31\;GeV^2.
\]

There are eight meson Regge trajectories populated by $u-d$ bound states
(for each isospin $I$ and angular momenta $J=l\pm 1$, $J=l$ and total spin $S
$ of $q\bar q$ system, $S=0,1$. Some of the trajectories are plotted in Fig.
2-4.

We observe that all experimental data are in good agreement with spectrum
given by Eq. (\ref{spect}) for $l\geq 1$ and $S=1$ (Fig.2) .
\begin{figure}[thb]
\vspace*{-0.2cm}
\epsfxsize = 8cm
\epsffile{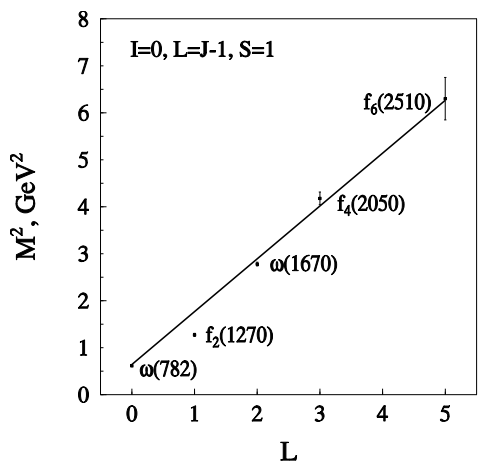}
\vspace*{-1.4cm}
\caption{$\omega$-meson Regge trajectory}
\end{figure}
As in the case of bound states with $S=0,$ the agreement between our
theoretical predictions and the existing experimental data is not good (see
Fig.3-4). Such deviations can be explained by absence of spin-dependent
terms and short-distance term of the potential.

\begin{figure}[thb]
\vspace*{-0.2cm}
\epsfxsize = 8cm
\epsffile{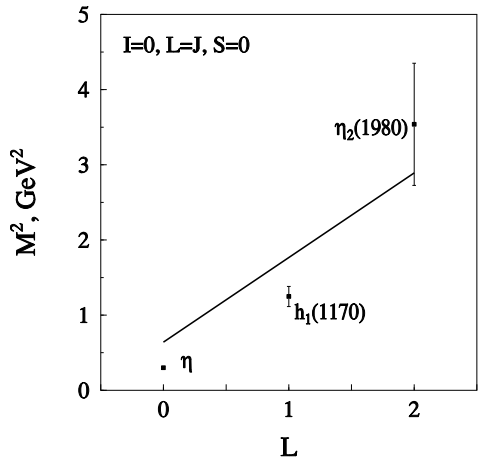}
\vspace*{-1.2cm}
\caption{$\eta$-meson Regge trajectory}
\end{figure}
\begin{figure}[thb]
\vspace*{-0.2cm}
\epsfxsize = 8cm
\epsffile{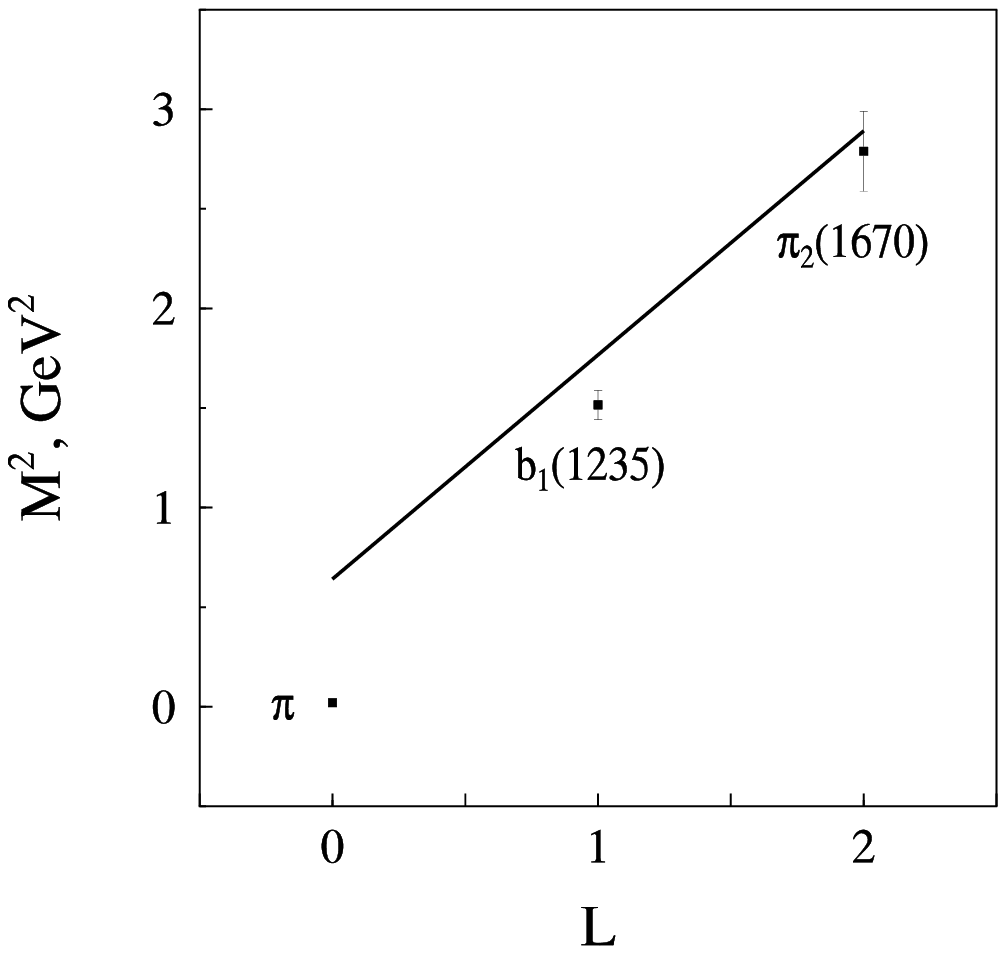}
\vspace*{-1.2cm}
\caption{$\pi$-meson Regge trajectory}
\end{figure}
As it is easy to see, our method for bound systems with different masses of
quarks does not require special processing of solving main equation of RQM,
as it was made in Ref.\cite{Kon1} (introduction of additional parameter).
When the masses of the quark and antiquark are different we obtain
\[
M^2 =2W_0+2\beta ^2\left( 2n+l+3/2\right) +m_1^2+m_2^2+
\]
\[
+ 2\sqrt{\left( W_0+2\beta ^2\left( 2n+l+3/2\right) \right)}*
\]
\begin{equation}
 *\sqrt{(W_0+2\beta^2
\left( 2n+l+3/2\right) +m_1^2+m_2^2)+m_1^2 m_2^2}
\label{spect1}
\end{equation}

We see that the dependence (\ref{spect1}) is, also, linear, but
asymptotically at large $l$. We apply (\ref{spect1}) to the description of
strange meson trajectories. For the mass of the $u(d)-$quark and for mass
strange quark we used the same value as in Ref. \cite{Godfrey}, i.e., $
m_1\equiv m_u=0.22\;GeV$ (see (\ref{um} ) , $m_2\equiv m_s=0.419\;GeV$. As
in the case of meson with hidden strangeness, the agreement between our
model predictions and existing experimental data \cite{Pdg1} is good (see
Fig.5).
\begin{figure}[thb]
\vspace*{-0.2cm}
\epsfxsize = 8cm
\epsffile{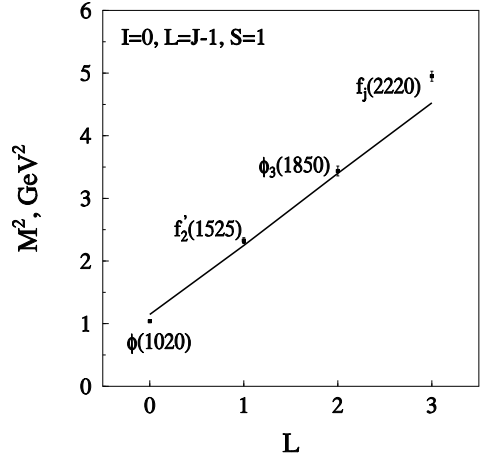}
\vspace*{-1.2cm}
\caption{$s-\bar s$-meson Regge trajectory}
\end{figure}

Also we can predict all the strange meson trajectories (see Fig.
6-9). The model describes all $s-\bar q (q-\bar s)$ meson Regge
trajectories with $S=1$ in quite satisfactory way (see Fig. 6-8).
But for mesons with $S=0$, the agreement between model and the
experimental data is not good (see Fig. 9), as in the case of $u-d$
mesons.
\begin{figure}[thb]
\vspace*{-0.2cm}
\epsfxsize = 8cm
\epsffile{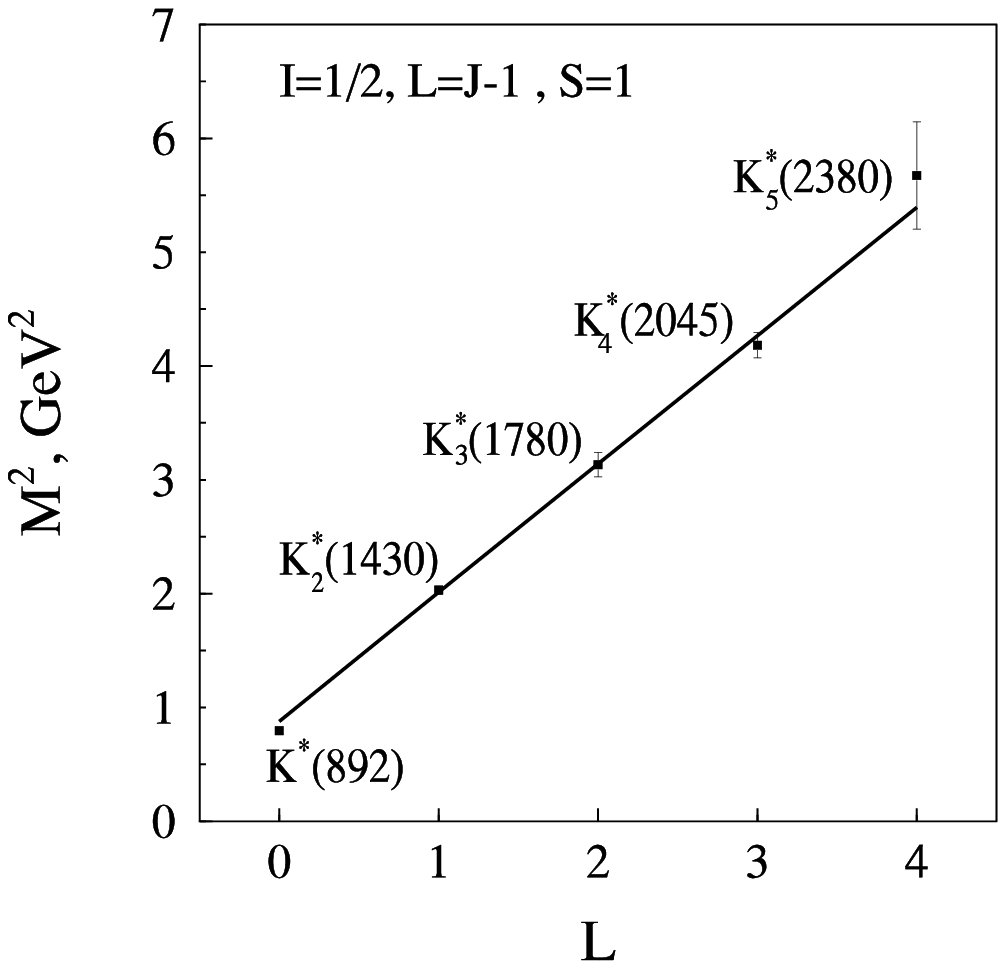}
\vspace*{-1.2cm}
\caption{$K$-meson Regge trajectory 1}
\end{figure}
\begin{figure}[thb]
\vspace*{-0.2cm}
\epsfxsize = 8cm
\epsffile{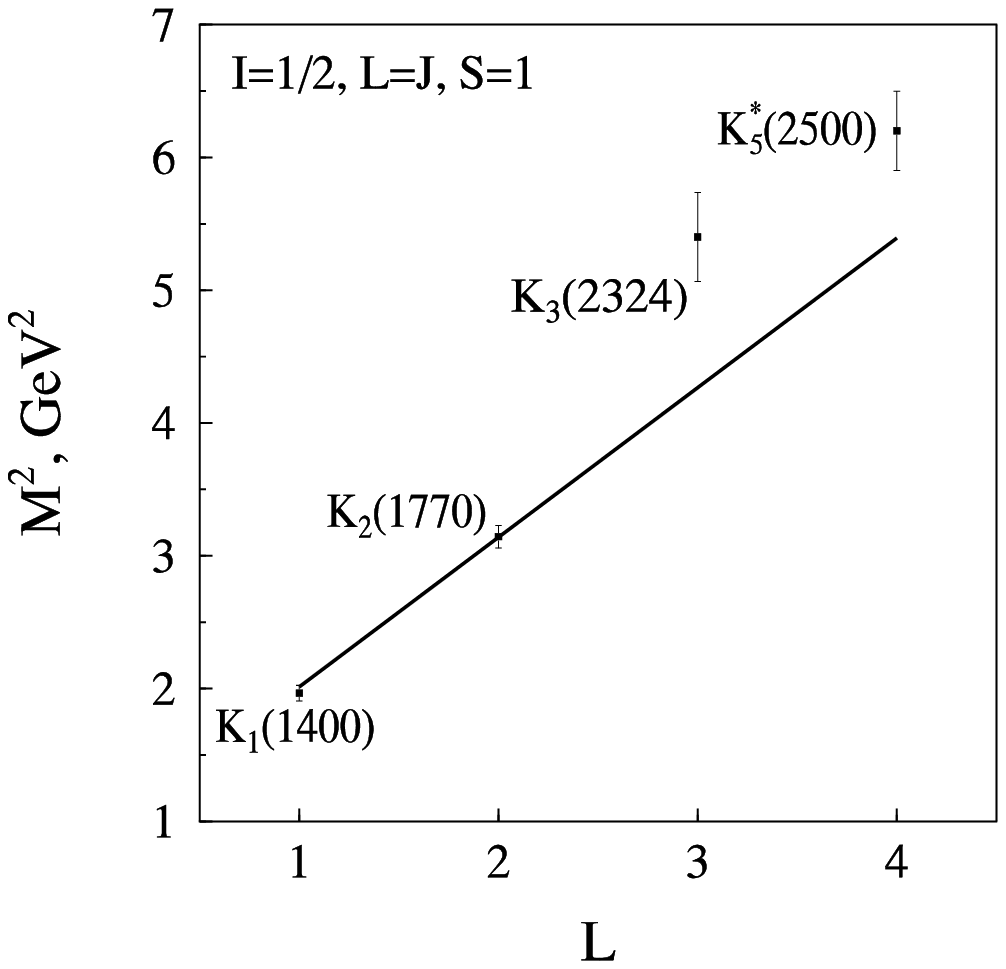}
\vspace*{-1.2cm}
\caption{$K$-meson Regge trajectory 2}
\end{figure}
\begin{figure}[thb]
\vspace*{-0.2cm}
\epsfxsize = 8cm
\epsffile{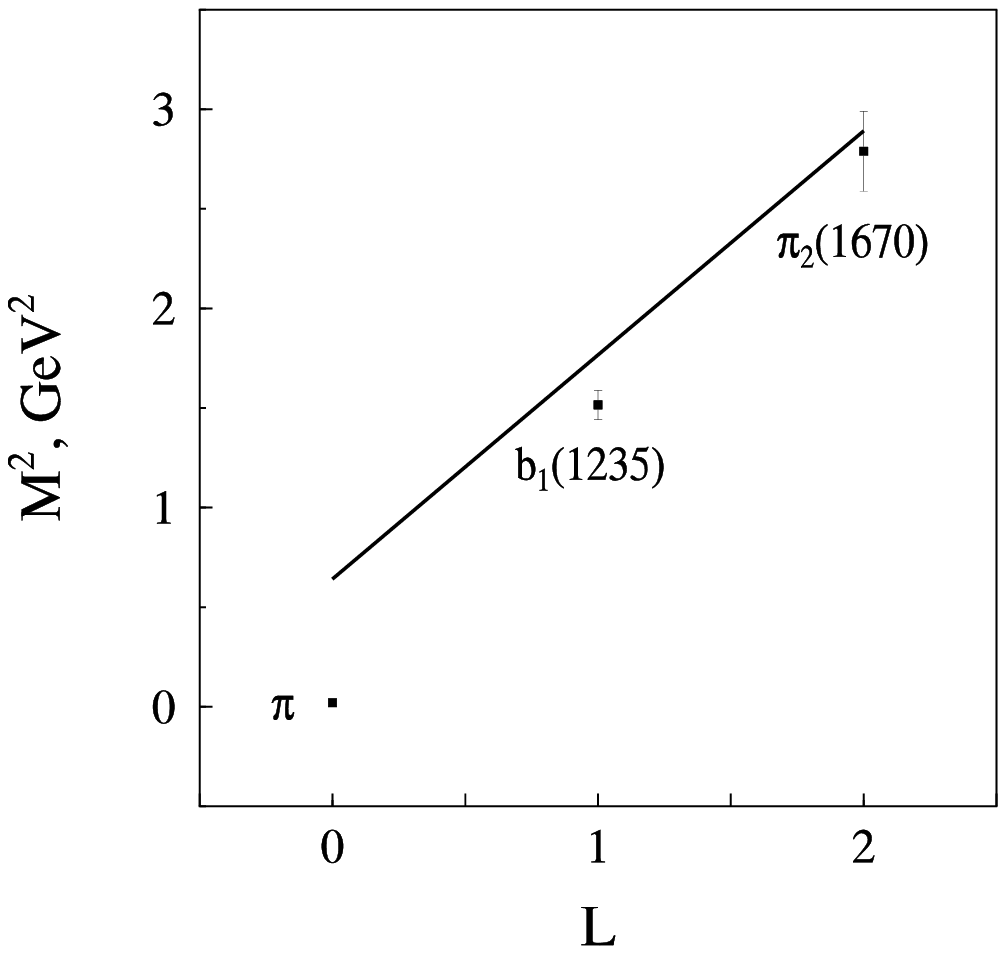}
\vspace*{-1.2cm}
\caption{$K$-meson Regge trajectory 3}
\end{figure}
\begin{figure}[thb]
\vspace*{-0.2cm}
\epsfxsize = 8cm
\epsffile{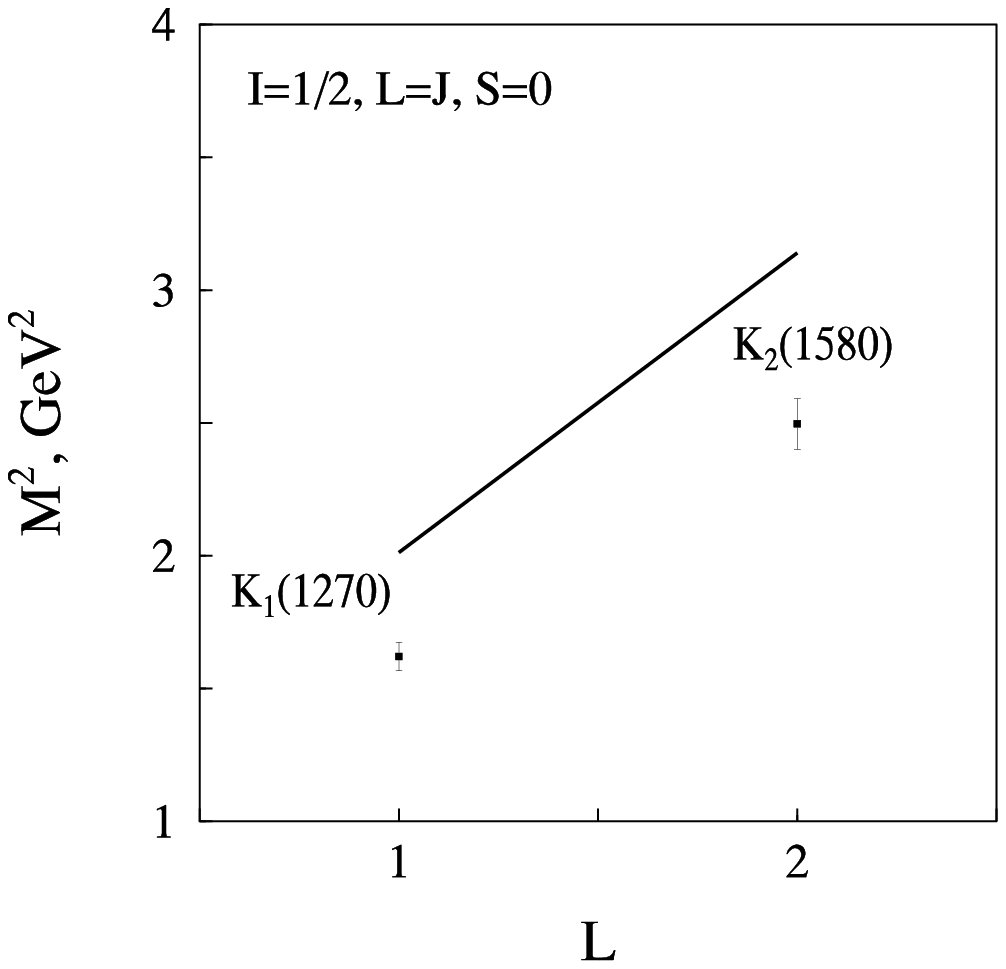}
\vspace*{-1.2cm}
\caption{$K$-meson Regge trajectory 4}
\end{figure}

In summary we shall mark that, the analytical solution of a main equation a
RQM (\ref{fun1}) can be used as zero approximation for problem solving with
more realistic potentials. The model adequately describes, with only four
free parameters, $m_u=m_d,m_s,W_0,\beta$, orbitally excited meson Regge
trajectories.

\end{document}